\documentclass[proceedings]{JHEP37}
\usepackage{eurosym}
\usepackage{amssymb}
\usepackage{amsfonts}
\usepackage{amsmath}
\usepackage{epsfig}

\newbox\mybox

\newcommand\fverb{\setbox\mybox=\hbox\bgroup\verb}
\newcommand\fverbdo{\egroup\medskip\noindent\fbox{\unhbox\mybox}\ }
\newcommand\fverbit{\egroup\item[\fbox{\unhbox\mybox}]}
\conference{Mending the broken PT-regime}
\abstract{We demonstrate that non-Hermitian Hamiltonian systems with spontaneously broken PT-symmetry and partially complex eigenvalue spectrum can be made meaningful 
in a quantum mechanical sense when introducing some explicit time-dependence into their parameters. Exploiting the fact that explicitly time-dependent non-Hermitian Hamitonians are unobservable and not 
identical to the energy operators in such a scenario, we show that their corresponding non-Hermitian energy operators develop a different type of PT-symmetry from the Hamiltonians 
that ensures the reality of their energy spectra. For this purpose we analytically solve the fully time-dependent Dyson equation with all quantities involved being explicitly time-dependent giving
rise to a time-dependent metric. The key auxiliary equation to be solved for the two level atomic system considered here is the nonlinear Ermakov-Pinney equation with time-dependent coefficients.}

\title{Mending the broken PT-regime via an explicit time-dependent Dyson map}
\author{Andreas Fring and Thomas Frith \\
Department of Mathematics, City, University of London,\\
Northampton Square, London EC1V 0HB, UK\\
E-mail: a.fring@city.ac.uk, thomas.frith@city.ac.uk}

\input{tcilatex}
\begin{document}

\section{Introduction}

It is well known that non-Hermitian Hamiltonians that commute with an
antilinear operator for which its eigenfunctions are eigenstates \cite{EW}
possess real eigenvalue spectra. $\mathcal{PT}$-symmetry \cite{Bender:1998ke}
is a specific example for such an antilinear symmetry for which many
examples have been worked out in detail, see e.g. \cite{BFGJ}. Moreover,
contrary to standard text book wisdom, such type of systems can be made
quantum mechanically meaningful \cite{Urubu,Benderrev,Alirev} by introducing
new inner products for which operators associated to observables are
self-adjoint. However, for systems with infinite dimensional Hilbert spaces
there are also well known issues related to the boundedness of the operators
involved \cite{siegl2012metric,bagarello2013self,FabFring1}. For instance,
while the metric operator might be bounded the inverse of the Dyson map,
needed to facilitate the mapping from a non-Hermitian Hamiltonian to an
isospectral Hermitian Hamiltonian, might be unbounded \cite{siegl2012metric}%
. There is also no guarantee that time-evolution operators for
time-independent non-Hermitian Hamiltonians with real eigenvalues are
bounded operators \cite{krejvcivrik2015pseudospectra}.

Another origin for the occurrence of unbounded time-evolution operators is
the spontaneously breaking of the $\mathcal{PT}$-symmetry. This scenario
emerges for $\mathcal{PT}$-symmetric Hamiltonians for which its eigenstates
are not eigenstates of the antilinear symmetry operator, in which case the
spectrum develops complex conjugate pairs of eigenvalues. While such a
situation is the most interesting one in optical settings \cite%
{Muss,MatMakris,Guo}, where different channels of gain and loss may be
constructed, such systems will inevitably develop infinite grows in energy
and are therefore usually discarded as being non-physical in a quantum
mechanical framework. We demonstrate here that by introducing an explicit
time-dependence into the parameters of the quantum Hamiltonian, such systems
can be made physically meaningful. This possibility exists since in a
quantum mechanical context those type of Hamiltonians are no longer
associated to the observable energy operator, as that operator acquires an
additional time-dependent correction term.

In order to find that correction term one needs to solve the time-dependent
Dyson relation for the Dyson map. So far only few explicit solutions to
these relations are known and progress has been made in various stages. The
simplest scenario is to assume that only the Hamiltonian is explicitly
dependent on time, but the Dyson map or the closely related metric operator
are kept time-independent \cite{CA,CArev}. More involved is to include the
time-dependence in the latter operator with a focus on finding solutions 
\cite{fringmoussa,fringmoussa2} without investigating the properties of the
corresponding wavefunctions of the time-dependent Schr\"{o}dinger equation.
In \cite{AndTom1,AndTom2} we studied the interesting possibility to keep the
non-Hermitian Hamiltonian time-independent with an explicit time-dependence
in the Dyson map. This allowed us to solve time-dependent Hermitian
Hamiltonian systems by transferring the time-dependence from the Hamiltonian
to the Dyson map or metric operators when discussing expectation values. The
corresponding solutions to the time-dependent Schr\"{o}dinger equation were
found to be entirely consistent for a quantum mechanical description.

Here we extend the previous analysis and consider a fully time-dependent
scenario for all quantities involved, that is the non-Hermitian Hamiltonian $%
H(t)$ together with its Hermitian counterpart $h(t)$ both being the defining
quantities in the time-dependent Schr\"{o}dinger equations 
\begin{equation}
h(t)\phi (t)=i\hbar \partial _{t}\phi (t),\qquad \text{and\qquad }H(t)\Psi
(t)=i\hbar \partial _{t}\Psi (t).  \label{TS}
\end{equation}%
The time-dependent invertible Dyson operator $\eta (t)$ relates the
solutions of these two equations by%
\begin{equation}
\phi (t)=\eta (t)\Psi (t),  \label{sol}
\end{equation}%
as well as the two Hamiltonians via the time-dependent Dyson relation 
\begin{equation}
h(t)=\eta (t)H(t)\eta ^{-1}(t)+i\hbar \partial _{t}\eta (t)\eta ^{-1}(t).
\label{hH}
\end{equation}%
Following the standard arguments of time-independent $\mathcal{PT}$%
-symmetric/quasi-Hermitian quantum mechanics \cite%
{Bender:1998ke,Benderrev,Alirev}, by asserting that observable operators $%
\mathcal{O}$ in the non-Hermitian system need to be related to a
self-adjoint operator $o(t)$ in the Hermitian system as $o(t)=\eta (t)%
\mathcal{O}(t)\eta ^{-1}(t)$, this leads to the curious fact that the
Hamiltonian $H(t)$, being defined as the operator satisfying the Schr\"{o}%
dinger equation, is not observable. This feature has led to a controversy 
\cite{time1,time6} questioning whether it is at all possible to formulate a
consistent fully time-dependent framework for non-Hermitian Hamiltonian
systems. The conundrum is easily solved by making a clear distinction
between the observable energy operator 
\begin{equation}
\tilde{H}(t)=\eta ^{-1}(t)h(t)\eta (t)=H(t)+i\hbar \eta ^{-1}(t)\partial
_{t}\eta (t),  \label{Henergy}
\end{equation}%
and the unobservable Hamiltonian $H(t)$ satisfying the time-dependent Schr%
\"{o}dinger equation. In what follows we set $\hbar =1$. In an adiabatic
approximation the energy spectrum of this operator may be dealt with
consistently at each instance of time \cite{time7}. In turn, since $H(t)$ is
not an observable operator this also means that its eigenvalues do not have
to be real at any instance in time. It is this latter fact that we exploit
to make sense of a non-Hermitian Hamiltonian with complex conjugate
eigenvalues as self-consistent quantum mechanical system.

\section{A two-level system with spontaneously broken PT-symmetry}

To illustrate our point we consider a simple two-level spin model described
by the non-Hermitian Hamiltonian 
\begin{equation}
H=-\frac{1}{2}\left[ \omega \mathbb{I}+\lambda \sigma _{z}+i\kappa \sigma
_{x}\right] ,
\end{equation}%
with $\sigma _{x}$, $\sigma _{y}$, $\sigma _{z}$ denoting the Pauli
matrices, $\mathbb{I}$ the identity matrix and $\omega $, $\lambda $, $%
\kappa \in \mathbb{R}$. The two eigenvalues and eigenvectors for this
Hamiltonian are simply%
\begin{equation}
E_{\pm }=-\frac{1}{2}\omega \pm \frac{1}{2}\sqrt{\lambda ^{2}-\kappa ^{2}}%
,\quad \text{and}\quad \varphi _{\pm }=\left( 
\begin{array}{c}
i(-\lambda \pm \sqrt{\lambda ^{2}-\kappa ^{2}}) \\ 
\kappa%
\end{array}%
\right) .  \label{EEE}
\end{equation}%
Using Wigner's argument \cite{EW,Bender:1998ke} the reality of the energy
spectrum for $\left\vert \lambda \right\vert >\left\vert \kappa \right\vert $
is easily explained by identifying an antilinear symmetry operator, denoted
here as $\mathcal{PT}$ , that commutes with the Hamiltonian and for which $%
\varphi _{\pm }$ are simultaneous eigenstates of $H$ and $\mathcal{PT}$ 
\begin{equation}
\left[ \mathcal{PT},H\right] =0,\qquad \text{and\qquad }\mathcal{PT}\varphi
_{\pm }=e^{i\phi }\varphi _{\pm },  \label{PTH}
\end{equation}%
with $\phi \in \mathbb{R}$. When $\left\vert \lambda \right\vert >\left\vert
\kappa \right\vert $ in our example the symmetry operator is easily
identified as $\mathcal{PT}=\tau \sigma _{z}$ with $\tau $ denoting complex
conjugation. When $\left\vert \lambda \right\vert <\left\vert \kappa
\right\vert $ the last relation in (\ref{PTH}) no longer holds and the
eigenvalues become complex conjugate to each other, a scenario usually
referred to as spontaneously broken $\mathcal{PT}$-symmetry. For the
parameter range of the latter situation this Hamiltonian would be regarded
as non-physical from a quantum mechanical point of view as it possesses
channels of infinite grows in energy, such that the corresponding time
evolution operators would be unbounded.

However, when one introduces an explicit time-dependence into the
Hamiltonian, $H\rightarrow H(t)$, it no longer plays the role of the
observable energy operator so that the complex eigenvalues do not constitute
any interpretational obstacle. For a meaningful physical picture one only
needs to guarantee now that the expectation values of $\tilde{H}(t)$, as
defined in (\ref{Henergy}), are real and instead identify a new $\widetilde{%
\mathcal{PT}}$-symmetry to be responsible for this property%
\begin{equation}
\left[ \widetilde{\mathcal{PT}},\tilde{H}\right] =0,\qquad \text{and\qquad }%
\widetilde{\mathcal{PT}}\tilde{\varphi}_{\pm }=e^{i\tilde{\phi}}\tilde{%
\varphi}_{\pm },  \label{24}
\end{equation}%
with $\tilde{\varphi}_{\pm }$ denoting the eigenvectors of $\tilde{H}$ and $%
\tilde{\phi}\in \mathbb{R}$. Notice that $\mathcal{PT}$ and $\widetilde{%
\mathcal{PT}}$ are only symbols here to denote different types of antilinear
operators, which however to not send $t$ to $-t$ as the time is only a real
parameter in this context.

Let us therefore introduce an explicit time-dependence into the parameters
of $H$, via $\lambda \rightarrow \alpha \kappa (t)$, $\kappa \rightarrow
\kappa (t)$, and solve this problem for the time-dependent Hamiltonian 
\begin{equation}
H(t)=-\frac{1}{2}\left[ \omega \mathbb{I}+\alpha \kappa (t)\sigma
_{z}+i\kappa (t)\sigma _{x}\right] .  \label{H}
\end{equation}%
To find the precise form of $\tilde{H}(t)$ we need to solve first equation (%
\ref{hH}) for the Dyson map $\eta (t)$. As discussed in \cite%
{AndTom1,AndTom2}, this is most easily achieved by pre-selecting some
concrete form for $h(t)$\footnote{%
Alternatively one may also solve the time-dependent quasi-Hermiticity
relation $H^{\dagger }\rho (t)-\rho (t)H=i\hbar \partial _{t}\rho (t)$ for
the metric operator $\rho (t)$ and subsequently determine $\eta (t)$ from $%
\rho (t):=\eta ^{\dagger }(t)\eta (t)$. However, as argued in \cite{AndTom2}%
, usually this turns out to be more difficult.}. For simplicity we take this
to be 
\begin{equation}
h(t)=-\frac{1}{2}\left[ \omega \mathbb{I}+\chi (t)\sigma _{z}\right] ,
\label{h}
\end{equation}%
with $\chi (t)$ being a general undetermined function of time. Taking $\eta
(t)$ to be of the most generic Hermitian form by using the notation%
\begin{equation}
\eta (t)=\frac{1}{2}\mathbb{[}\eta _{1}(t)+\eta _{4}(t)]\mathbb{I+}\eta
_{2}(t)\sigma _{x}+\eta _{3}(t)\sigma _{y}\mathbb{+}\frac{1}{2}\mathbb{[}%
\eta _{1}(t)-\eta _{4}(t)]\sigma _{z},  \label{Ansatzeta}
\end{equation}%
with real functions $\eta _{i}(t)$, the time-dependent Dyson equation (\ref%
{hH}) for (\ref{H}) and (\ref{h}) is solved when the component functions of $%
\eta (t)$ satisfy the coupled first order equations 
\begin{eqnarray}
\dot{\eta}_{1} &=&\frac{\kappa }{2}\eta _{2},~~~~\dot{\eta}_{2}=\frac{\chi
+\alpha \kappa }{2}\eta _{3}+\frac{\kappa }{2}\eta _{1},~~~~\dot{\eta}_{3}=-%
\frac{\chi +\alpha \kappa }{2}\eta _{2},~~~~\dot{\eta}_{4}=\frac{\kappa }{2}%
\eta _{2},~~  \label{coupled} \\
\eta _{1} &=&\eta _{4},~~~~\chi =\kappa \left( \frac{\eta _{3}}{\eta _{1}}%
+\alpha \right) .
\end{eqnarray}%
The overdot denotes here as usual a differentiation with respect to time.
The equations (\ref{coupled}) are solved by%
\begin{equation}
\eta _{1}=\eta _{4}=c\sqrt{\frac{\kappa }{\chi }},\qquad \eta _{2}=\frac{c}{%
\sqrt{\kappa \chi }}\left( \frac{\dot{\kappa}}{\kappa }-\frac{\dot{\chi}}{%
\chi }\right) ,\qquad \eta _{3}=c\left( \sqrt{\frac{\chi }{\kappa }}-\alpha 
\sqrt{\frac{\kappa }{\chi }}\right) ,  \label{eta}
\end{equation}%
with $c$ denoting an integration constant and $\chi (t)$ satisfying the
nonlinear second order equation%
\begin{equation}
\ddot{\chi}-\frac{3}{2}\frac{\dot{\chi}^{2}}{\chi }+\left[ \frac{3}{2}\left( 
\frac{\dot{\kappa}}{\kappa }\right) ^{2}-\frac{\ddot{\kappa}}{\kappa }+\frac{%
1}{2}\kappa ^{2}(1-\alpha ^{2})\right] \chi +\frac{\chi ^{3}}{2}=0.
\label{xi}
\end{equation}%
Using the parameterizations $\chi =2/\sigma ^{2}$ or $\kappa =2/(\sigma ^{2}%
\sqrt{\alpha ^{2}-1})$ this equation is converted into the Ermakov-Pinney
(EP) equation \cite{Ermakov,Pinney} for $\sigma $ 
\begin{equation}
\ddot{\sigma}+\lambda (t)\sigma =\frac{1}{\sigma ^{3}}  \label{EPtime}
\end{equation}%
with time-dependent coefficient%
\begin{equation}
\lambda (t)=\frac{1}{2}\frac{\ddot{\kappa}}{\kappa }-\frac{3}{4}\left( \frac{%
\dot{\kappa}}{\kappa }\right) ^{2}-\frac{1}{4}\kappa ^{2}(1-\alpha
^{2})~~~~~~\text{or~~~~~}\lambda (t)=\frac{1}{2}\frac{\ddot{\chi}}{\chi }-%
\frac{3}{4}\left( \frac{\dot{\chi}}{\chi }\right) ^{2}+\frac{1}{4}\chi ^{2},
\label{lambda}
\end{equation}%
respectively. Thus either way given the time-dependent field $\kappa (t)$ in 
$H(t)$ or $\chi (t)$ in $h(t)$ the remaining field is constrained by the EP
equation with almost identical coefficients. The EP equation emerges in many
scenarios of time-dependent quantum mechanics and various areas in
mathematics, see for instance \cite{leach2008ermakov} for an overview. The
general solution for (\ref{EPtime}), as reported by Pinney \cite{Pinney}, is%
\begin{equation}
\sigma (t)=\left( Au^{2}+Bv^{2}+2Cuv\right) ^{1/2},  \label{EPsol}
\end{equation}%
where $u(t)$ and $v(t)$ are the two fundamental solutions to the equation $%
\ddot{\sigma}+\lambda (t)\sigma =0$ and the constants $A$, $B$, $C$ are
constrained as $C^{2}=AB-W^{-2}$ with $W=u\dot{v}-v\dot{u}$ denoting the
corresponding Wronskian. Thus from the solution of the EP equation for fixed 
$\alpha $ we can obtain now a specific solution for the Dyson map (\ref{eta}%
). As the exceptional point at $\alpha =1$ for $H$ leads to qualitatively
different solutions, we treat this case separately from the cases with $%
\alpha \neq 1$.

\subsection{The $\protect\widetilde{\mathcal{PT}}$-symmetric regimes of $%
\tilde{H}$, $\protect\alpha \neq 1$}

We are left with solving the EP equation so that (\ref{coupled}) becomes an
explicit solution to the time-dependent Dyson equation. For definiteness we
assume here that $\kappa (t)$ is given and determine $\chi (t)$, but as
mentioned in the previous section the reverse computation requires very
little modification. Taking the time-dependent coefficient $\lambda (t)$ in
the EP equation to be of the form (\ref{lambda}) for $\alpha \neq 1$ we find%
\begin{equation}
u(t)=\frac{1}{\sqrt{\kappa }}e^{\mu /2},\quad v(t)=\frac{1}{\sqrt{\kappa }}%
e^{-\mu /2},\quad \text{with }\mu (t):=\sqrt{1-\alpha ^{2}}%
\int\nolimits^{t}\kappa (s)ds,
\end{equation}%
such that the solution to the EP equation (\ref{EPsol}) becomes 
\begin{equation}
\sigma (t)=\frac{1}{\sqrt{\kappa }}\left( Ae^{\mu }+Be^{-\mu }\pm 2\sqrt{%
AB-1/(1-\alpha ^{2})}\right) ^{1/2}.\quad ~~
\end{equation}%
Parameterizing the constants further as $A=c_{1}+c_{2}$, $B=c_{1}-c_{2}$ we
obtain the solution 
\begin{equation}
\chi (t)=\frac{\kappa }{\xi },~\ \ \text{with }\xi :=c_{1}\cosh \mu
+c_{2}\sinh \mu \pm \sqrt{c_{1}^{2}-c_{2}^{2}-1/(1-\alpha ^{2})}.
\end{equation}%
Thus the Dyson map is obtained from (\ref{eta}) as%
\begin{equation}
\eta _{1}=\eta _{4}=\sqrt{\xi },~~~\eta _{2}=\frac{\hat{\xi}\sqrt{1-\alpha
^{2}}}{\sqrt{\xi }},~~~\eta _{3}=\frac{1-\alpha \xi }{\sqrt{\xi }},~~~~\text{%
with }\hat{\xi}:=c_{1}\sinh \mu +c_{2}\cosh \mu
\end{equation}%
Since in all relevant equations $\eta $ is accompanied by it inverse we have
set $c=1$ in (\ref{eta}) without loss of generality. Noting that $\det \eta
=\eta _{1}^{2}-\eta _{2}^{2}-\eta _{3}^{2}=\pm 2\delta $ with $\delta
:=\alpha +(1-\alpha ^{2})\sqrt{c_{2}^{2}-c_{3}^{2}-1/(1-\alpha ^{2})}$ the
Dyson map is invertible for as long as $\alpha \neq 0$ or $c_{2}^{2}\neq
c_{3}^{2}+1/(1-\alpha ^{2})$.

Next we turn to solving the time-dependent Schr\"{o}dinger equation. This is
easily achieved for the first equation in (\ref{TS}) as $h(t)$ is diagonal.
We find the two orthonormal solutions 
\begin{equation}
\left\vert \phi _{+}(t)\right\rangle =e^{i\omega t/2+i\theta (t)}\left( 
\begin{array}{c}
1 \\ 
0%
\end{array}%
\right) \qquad \text{and\qquad }\left\vert \phi _{-}(t)\right\rangle
=e^{i\omega t/2-i\theta (t)}\left( 
\begin{array}{c}
0 \\ 
1%
\end{array}%
\right) ,  \label{ef}
\end{equation}%
with $\left\langle \phi _{i}(t)\left\vert \phi _{j}(t)\right\rangle \right.
=\delta _{ij}$ for $i,j=+,-$ and 
\begin{equation}
\theta (t)=\frac{1}{2}\int\nolimits^{t}\chi (s)ds=\arctan \left\{ \sqrt{%
1-\alpha ^{2}}\left[ c_{2}+\left( c_{1}\mp \sqrt{c_{1}^{2}-c_{2}^{2}-\frac{1%
}{1-\alpha ^{2}}}\right) \tanh \left[ \mu (t)/2\right] \right] \right\} .
\end{equation}%
Having found $\eta (t)$ we obtain from (\ref{sol}) the solution for the Schr%
\"{o}dinger equation related to the non-Hermitian Hamiltonian $H(t)$ as%
\begin{equation}
\left\vert \psi _{+}(t)\right\rangle =\frac{-e^{i\omega t/2+i\theta (t)}}{%
2\delta }\left( 
\begin{array}{c}
-\eta _{1} \\ 
\eta _{2}+i\eta _{3}%
\end{array}%
\right) ~~\text{and~~}\left\vert \psi _{-}(t)\right\rangle =\frac{e^{i\omega
t/2-i\theta (t)/}}{2\delta }\left( 
\begin{array}{c}
\eta _{2}-i\eta _{3} \\ 
-\eta _{1}%
\end{array}%
\right) .  \label{EH}
\end{equation}%
By construction these states are orthonormal with regard to the inner
product with modified metric $\left\langle \psi _{i}(t)\left\vert \eta
^{2}\psi _{j}(t)\right\rangle \right. =\delta _{ij}$ for $i,j=+,-$. Next we
compute the energy operator (\ref{Henergy}), which acquires the form%
\begin{equation}
\tilde{H}(t)=-\frac{1}{2}\left\{ \omega \mathbb{I}+\frac{\chi }{\delta }%
\left[ i\left( \alpha \xi -1\right) \sigma _{x}+i\left( \hat{\xi}\sqrt{%
1-\alpha ^{2}}\right) \sigma _{y}+(\xi -\delta )\sigma _{z}\right] \right\} .
\end{equation}%
Since $\tilde{H}(t)$ is related to a Hermitian Hamiltonian by a similarity
transformation we expect the eigenvalues of this Hamiltonian to be real when
this transformation is well defined. Indeed, it turns out that the energy
expectation values for these states are real at any instance in time and
simply result to%
\begin{equation}
\tilde{E}_{\pm }(t)=\left\langle \psi _{\pm }(t)\left\vert \tilde{H}(t)\eta
^{2}\psi _{\pm }(t)\right\rangle \right. =\left\langle \phi _{\pm
}(t)\left\vert h(t)\phi _{\pm }(t)\right\rangle \right. =-\frac{1}{2}\left[
\omega \pm \chi (t)\right] .  \label{ETT}
\end{equation}%
Thus as long as $\chi (t)$ is real the energy expectation values are real,
which is the case when $c_{1}$,$c_{2}\in \mathbb{R}$ and $%
c_{1}^{2}>c_{2}^{2}+1/(1-\alpha ^{2})$ for $\alpha <1$ or when $c_{1}\in 
\mathbb{R}$ and $c_{2}\in i\mathbb{R}$ for $\alpha >1$. We depict the energy
spectra as a function of time for some specific parameter values in figures %
\ref{energy} and \ref{energy2}.

\FIGURE{ \epsfig{file=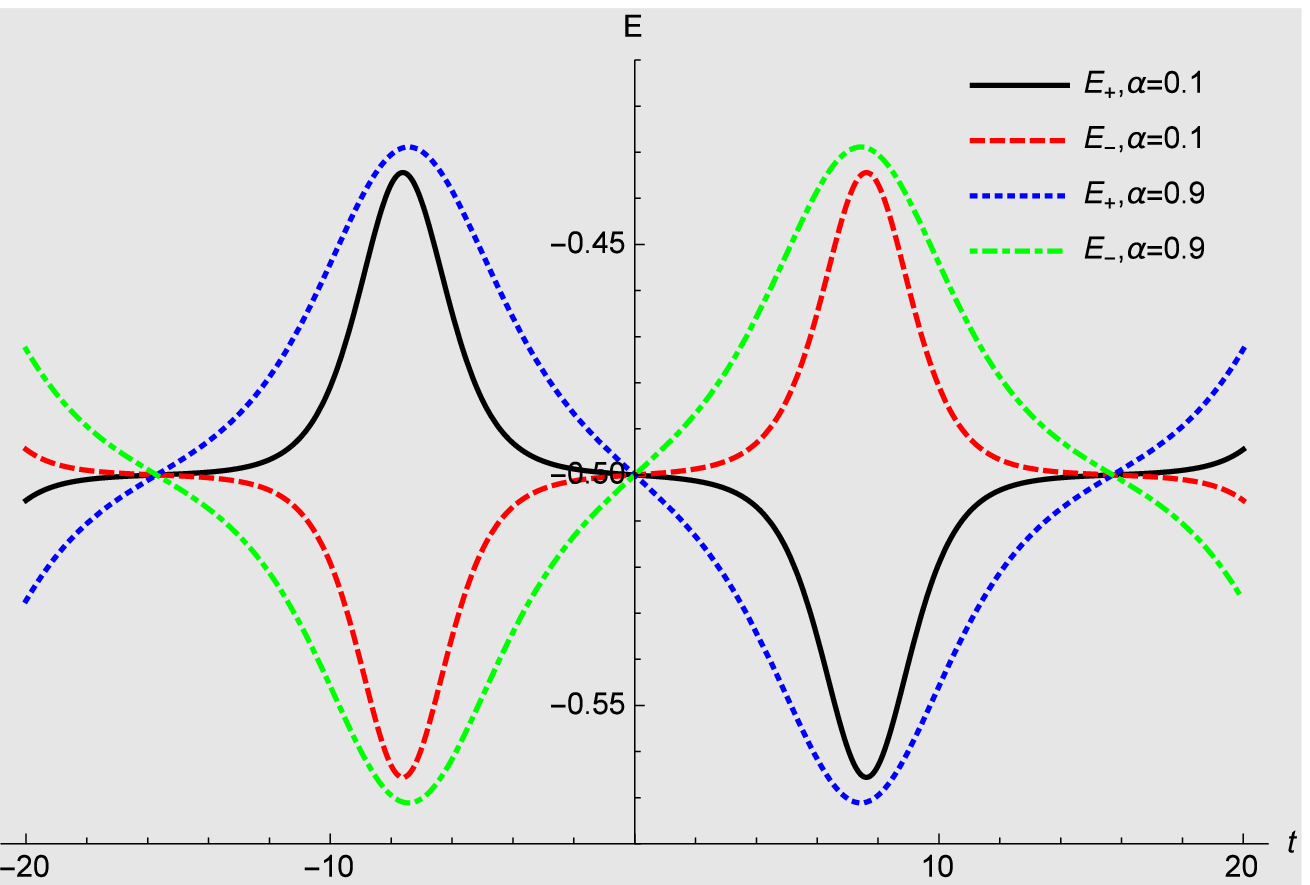,width=6.9cm} \epsfig{file=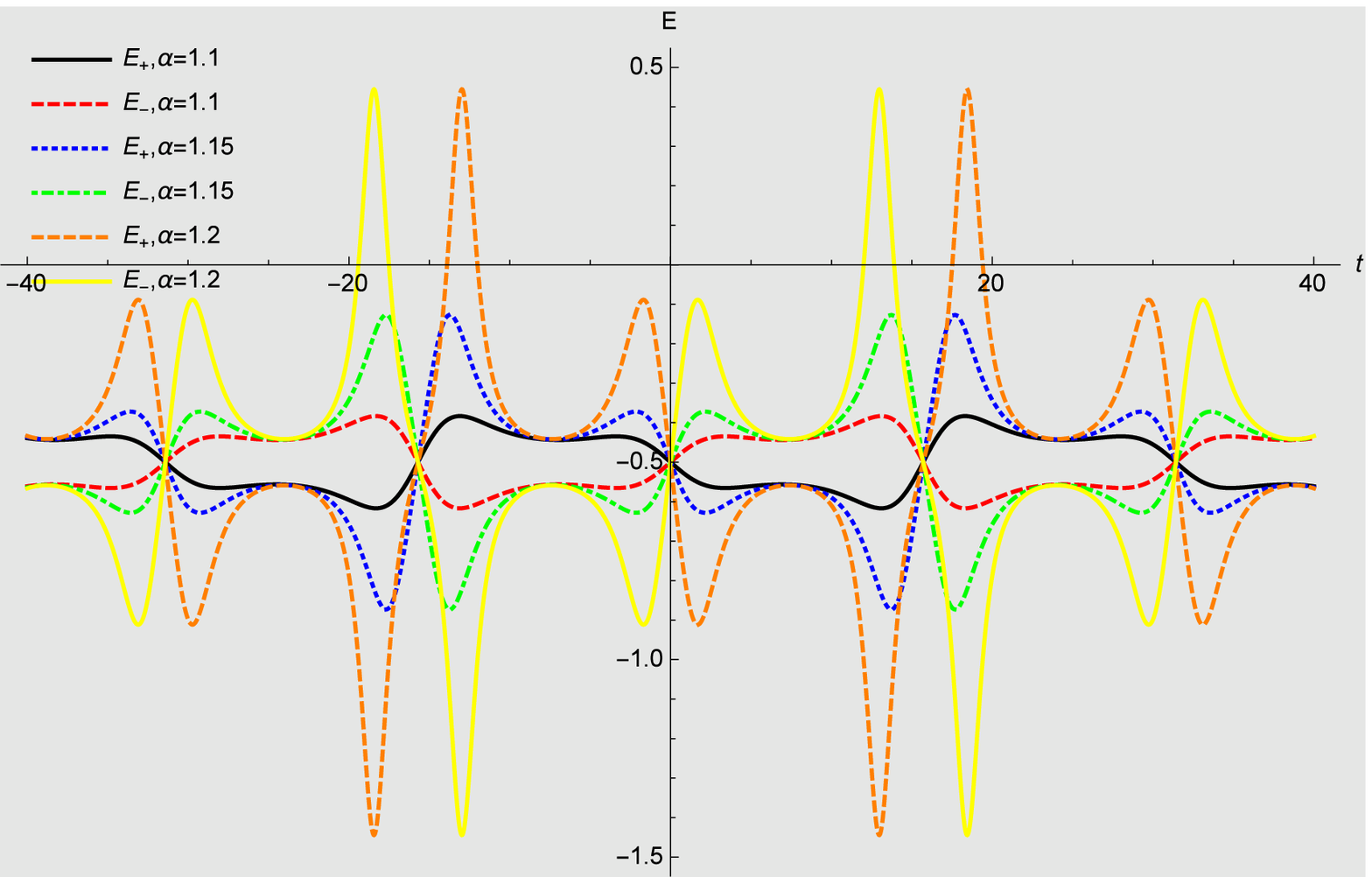,width=7.2cm}
\caption{Real energies in the $\widetilde{\mathcal{PT}}$-symmetric phase for $\kappa(t)=\sin(t/5)$, $\omega=1$, $c_1=4$ and $c_2=1$ (panel a)  $c_2=i$ (panel b).}
        \label{energy}}

The behaviour shown in the figures \ref{energy} and \ref{energy2} is typical
for non-Hermitian with an antilinear symmetry. Thus we expect for $\tilde{H}%
(t)$ that there exists an antilinear symmetry $\widetilde{\mathcal{PT}}$
that solves (\ref{24}) and hence explains the reality and complexity of the
eigenspectrum. Evidently the operator $\mathcal{PT}$ as introduced above is
not the correct symmetry and only serves to explain the spectrum for $H(t)$.
Thus we make a generic Ansatz for this operator and try to solve the first
relation in (\ref{24}). Indeed we find as the unique solution the antilinear
operator

\FIGURE{ \epsfig{file=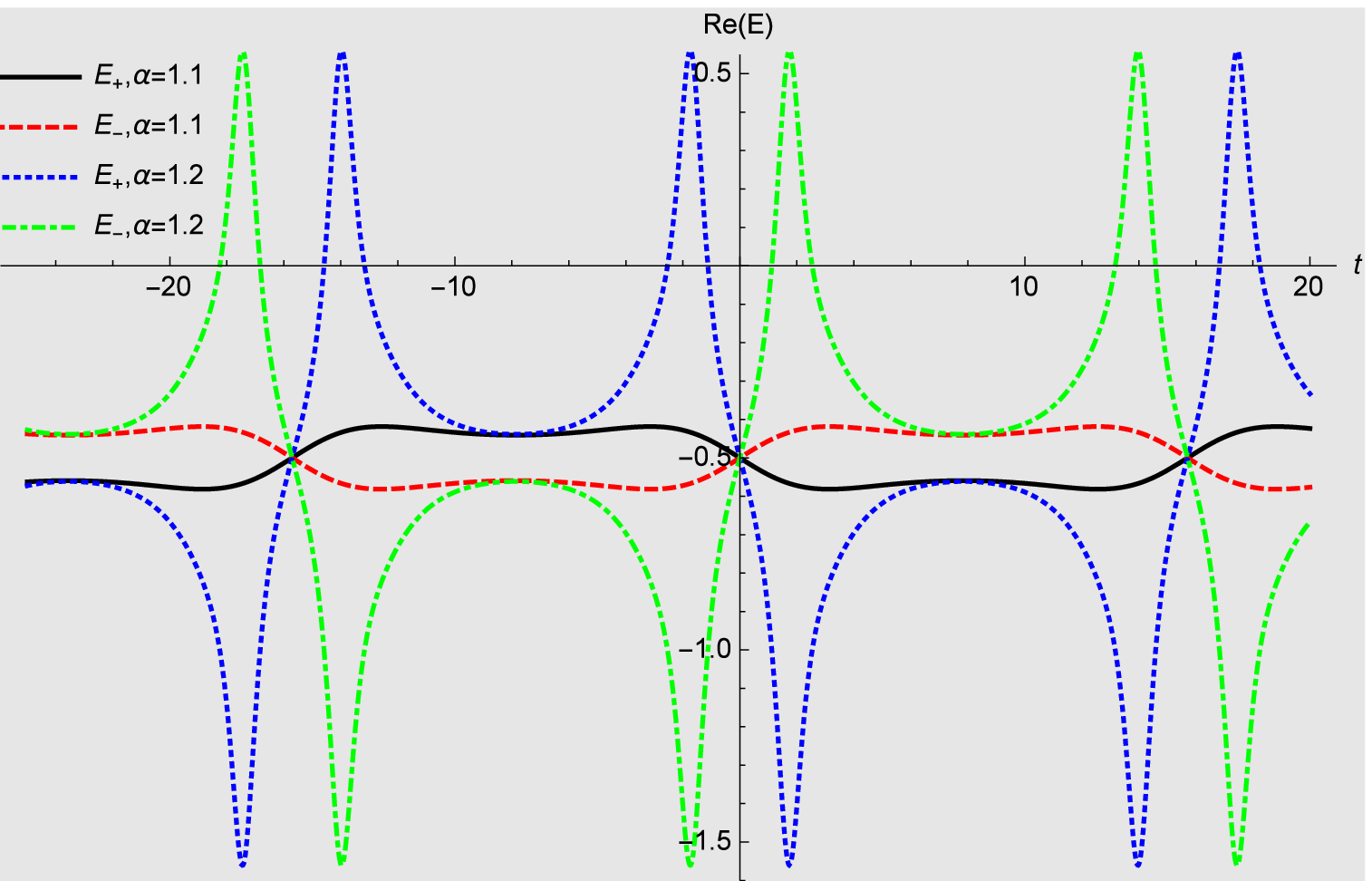,width=7.0cm} \epsfig{file=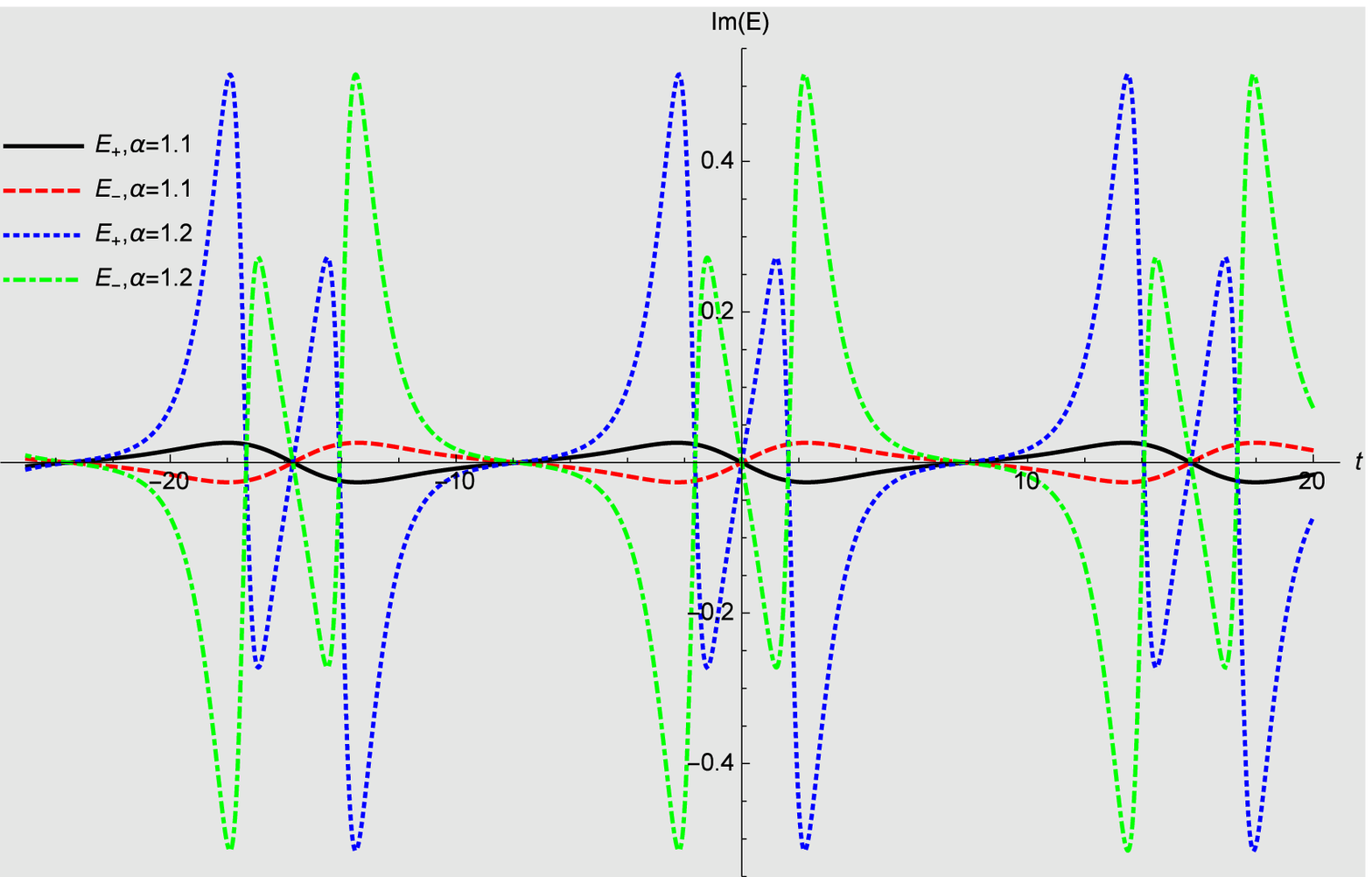,width=7.0cm} \epsfig{file=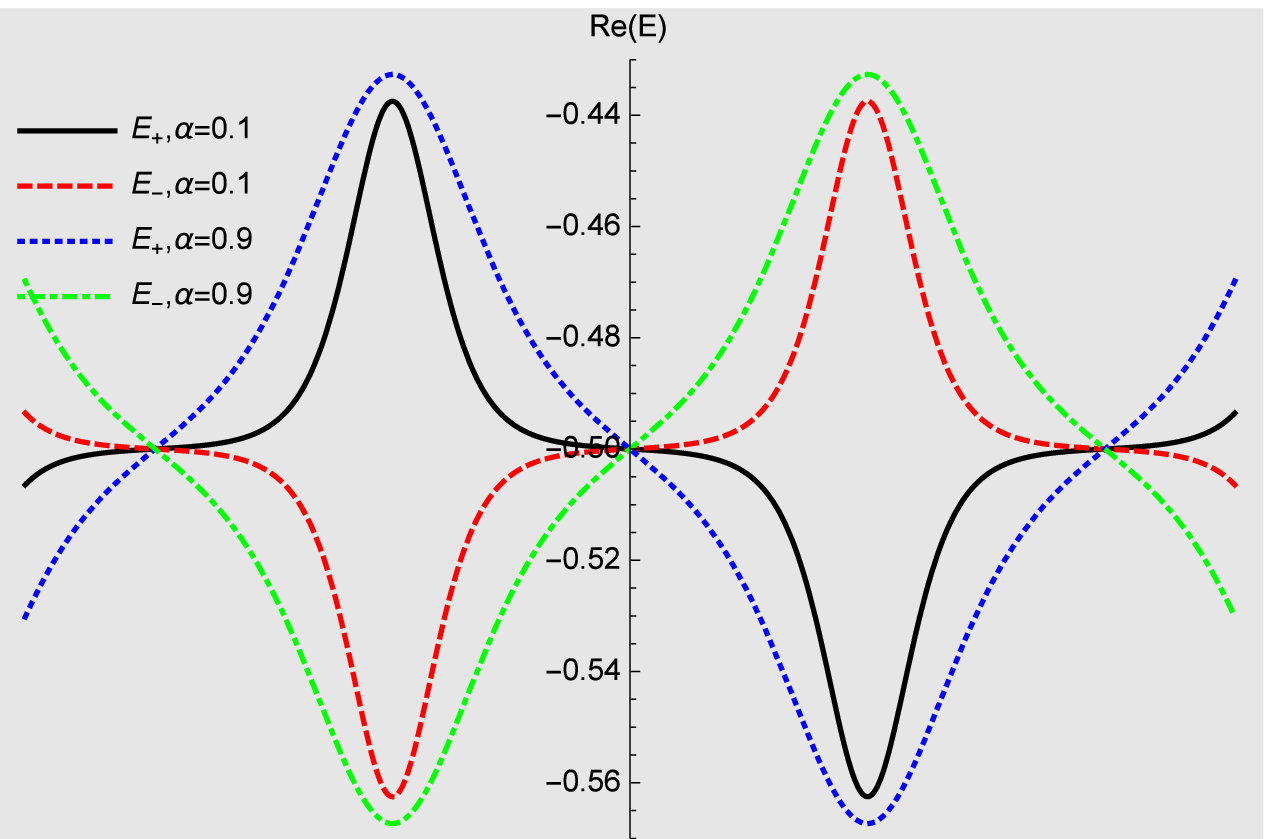,width=7.0cm} \epsfig{file=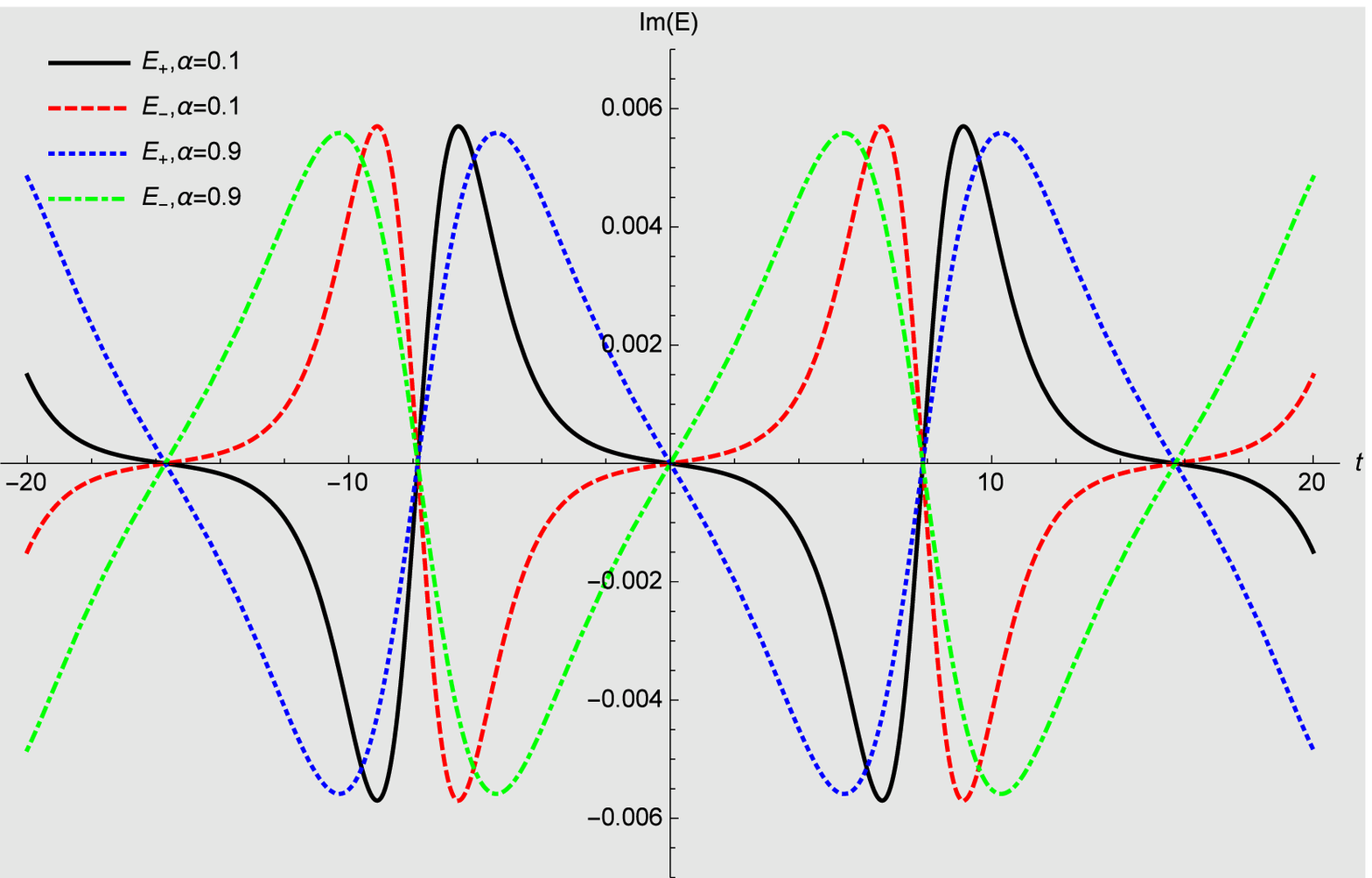,width=7.0cm}
\caption{Complex energies in the $\widetilde{\mathcal{PT}}$-broken phase for $\kappa(t)=\sin(t/5)$, $\omega=1$, $c_1=4$ and $c_2=1$ (panel a, b) $c_2=i$ (panel c, d).}
        \label{energy2}}

\begin{equation}
\widetilde{\mathcal{PT}}:=\frac{1}{\sqrt{(\xi -\delta )^{2}+(\alpha ^{2}-1)%
\hat{\xi}^{2}}}\left[ i\left( \sqrt{1-\alpha ^{2}}\hat{\xi}\right) \sigma
_{y}+(\xi -\delta )\sigma _{z}\right] \tau .  \label{PT}
\end{equation}%
We verify that $\widetilde{\mathcal{PT}}$ is involutory with $\widetilde{%
\mathcal{PT}}^{2}=\mathbb{I}$. Furthermore we verify that $\widetilde{%
\mathcal{PT}}\sigma _{x}\widetilde{\mathcal{PT}}=-\sigma ^{x}$ and $%
\widetilde{\mathcal{PT}}\sigma _{z}\widetilde{\mathcal{PT}}\neq \sigma _{z}$%
. Thus when $\alpha \neq 0$ the new $\widetilde{\mathcal{PT}}$-symmetry is
not a symmetry of $H(t)$, i.e. we have $\left[ \widetilde{\mathcal{PT}},H(t)%
\right] \neq 0$ but $\left[ \widetilde{\mathcal{PT}},\tilde{H}(t)\right] =0$%
. In order to guarantee that this symmetry is unbroken we also need to
satisfy the second equation in (\ref{24}). We determine the eigenvectors of $%
\tilde{H}(t)$ as%
\begin{equation}
\tilde{\varphi}_{\pm }\sim \left( 
\begin{array}{c}
(1\mp 1)\delta -\xi \\ 
\sqrt{1-\alpha ^{2}}\hat{\xi}+i(1-\alpha \xi )%
\end{array}%
\right) ,
\end{equation}%
and verify that these vectors are indeed $\widetilde{\mathcal{PT}}$%
-eigenstates 
\begin{equation}
\widetilde{\mathcal{PT}}\tilde{\varphi}_{\pm }=e^{i\tilde{\omega}_{\pm }}%
\tilde{\varphi}_{\pm }.
\end{equation}%
with 
\begin{eqnarray}
\tilde{\omega}_{+} &=&\arctan \left[ \frac{2\sqrt{1-\alpha ^{2}}(1-\alpha
\xi )\hat{\xi}}{1+\xi (\xi -2\alpha +\xi \alpha ^{2})+(\alpha ^{2}-1)\hat{\xi%
}^{2}}\right] , \\
\tilde{\omega}_{-} &=&\arctan \left[ \frac{\sqrt{1-\alpha ^{2}}(1-\alpha \xi
)\hat{\xi}}{2\delta ^{2}-3\delta \xi +\xi ^{2}+(\alpha ^{2}-1)\hat{\xi}^{2}}%
\right] +\pi .
\end{eqnarray}%
Thus for the regime stated above the $\widetilde{\mathcal{PT}}$-symmetry is
unbroken and the eigenvalues of $\tilde{H}(t)$ are therefore guaranteed to
be real. We notice that for $a>1$ also the Hamiltonian $H(t)$ is in its $%
\mathcal{PT}$ -symmetric phase, but $\widetilde{\mathcal{PT}}$ is still not
a symmetry for $H(t)$.

\subsection{The $\protect\widetilde{\mathcal{PT}}$-symmetric regime of $%
\tilde{H}$ and $\protect\widetilde{\mathcal{PT}}$-broken regime of $H$, $%
\protect\alpha =0$}

The value $\alpha =0$ is special as in this case the $\widetilde{\mathcal{PT}%
}$-operator commutes with both $\tilde{H}(t)$ and $H(t)$, but the
eigenvalues of the latter (\ref{EEE}) are complex conjugate in this case.
This means we expect the eigenvectors of $H(t)$ not to be eigenstates of the 
$\widetilde{\mathcal{PT}}$-operator. It is instructive to verify this in
detail and since the formulae simplify substantially in this case, it is
also useful to have a simpler example at hand. The two orthonormal solutions
for $h(t)$ take on the same form as in (\ref{ef}) with 
\begin{equation}
\theta (t)=\frac{1}{2}\int\nolimits^{t}\chi (s)ds=\arctan \left\{
c_{2}+\left( c_{1}\mp \sqrt{c_{1}^{2}-c_{2}^{2}-1}\tanh \left[ \mu (t)/2%
\right] \right) \right\} ,
\end{equation}%
and the solutions for the Schr\"{o}dinger equation related to the
non-Hermitian Hamiltonian $H(t)$ are also given by (\ref{EH}) with $\det
\eta =2\sqrt{c_{2}^{2}-c_{3}^{2}-1}$. The energy operator (\ref{Henergy})
simplifies to%
\begin{equation}
\tilde{H}(t)=-\frac{1}{2}\omega \mathbb{I}+\frac{\chi (t)}{\sqrt{%
c_{1}^{2}-c_{2}^{2}-1}}\left[ i\sigma _{x}-(c_{1}\sinh \mu +c_{2}\cosh \mu
)i\sigma _{y}-(c_{1}\cosh \mu +c_{2}\sinh \mu )\sigma _{z}\right] .
\end{equation}%
and the $\widetilde{\mathcal{PT}}$-operator reduces to%
\begin{equation}
\widetilde{\mathcal{PT}}:=\frac{1}{\sqrt{c_{1}^{2}-c_{2}^{2}}}\left[
(c_{1}\sinh \mu +c_{2}\cosh \mu )i\sigma _{y}+(c_{1}\cosh \mu +c_{2}\sinh
\mu )\sigma _{z}\right] \tau .
\end{equation}%
Now both Hamiltonians are $\widetilde{\mathcal{PT}}$-symmetric, i.e. in
addition to $\left[ \widetilde{\mathcal{PT}},\tilde{H}(t)\right] =0$ we also
have $\left[ \widetilde{\mathcal{PT}},H(t)\right] =0$. However, whereas the
eigenvectors $\tilde{\varphi}_{+}\sim \{-\eta _{1},\eta _{2}+i\eta _{3}\}$, $%
\tilde{\varphi}_{-}\sim \{\eta _{2}-i\eta _{3},\eta ,\}$ of $\tilde{H}(t)$
are $\widetilde{\mathcal{PT}}$-symmetric, the eigenvectors $\varphi _{\pm
}\sim \{\pm 1,1\}$ of $H(t)$ are not eigenstates of the $\widetilde{\mathcal{%
PT}}$-operator. Hence we have 
\begin{equation}
\widetilde{\mathcal{PT}}\varphi _{\pm }\neq e^{i\omega _{\pm }}\varphi _{\pm
}\qquad \text{and\qquad }\widetilde{\mathcal{PT}}\tilde{\varphi}_{\pm }=e^{i%
\tilde{\omega}_{\pm }}\tilde{\varphi}_{\pm }.
\end{equation}%
Concretely we identity 
\begin{equation}
\tilde{\omega}_{\pm }=\arctan \left[ \pm \frac{c_{2}^{2}-c_{1}^{2}-(c_{1}%
\cosh \mu +c_{2}\sinh \mu )\sqrt{c_{1}^{2}-c_{2}^{2}-1}}{c_{1}\sinh \mu
+c_{2}\cosh \mu }\right] .
\end{equation}%
Thus the $H(t)$ system is always in the spontaneously broken $\widetilde{%
\mathcal{PT}}$-symmetry phase whereas $\tilde{H}(t)$ is $\widetilde{\mathcal{%
PT}}$-symmetric as long as $c_{1}^{2}-c_{2}^{2}>1$.

\subsection{The exceptional point of $H(t)$ at $\protect\alpha =1$}

The value $\alpha =1$ is an exceptional point for $H(t)$ as it marks the
transition from real to complex conjugate eigenvalues and at the same time
the two eigenvectors coalesce. For $\tilde{H}$ it also indicates the
boundary of the real eigenvalues, but they do not become complex conjugate
to each other and the two eigenvectors remain different. The EP equation
admits a qualitatively different solution in this case. Taking the
time-dependent coefficient to be of the form (\ref{lambda}) for $\alpha =1$
with given $\kappa $ we find%
\begin{equation}
u(t)=\frac{1}{\sqrt{\kappa }},\quad v(t)=\frac{1}{\sqrt{\kappa }}\mu ,\quad 
\text{with }\mu :=\int\nolimits^{t}\kappa (s)ds,
\end{equation}%
such that the solution to the EP equation (\ref{EPsol}) becomes 
\begin{equation}
\sigma (t)=\sqrt{\frac{\mu }{\kappa }}\left( A\mu +B\mu ^{-1}\pm 2\sqrt{AB-1}%
\right) ^{1/2}.\quad ~~
\end{equation}%
so that 
\begin{equation}
\chi (t)=\frac{\kappa }{\xi },~\ \ \text{with }\xi :=\frac{1}{2}\left(
B+A\mu ^{2}\pm 2\mu \sqrt{AB-1}\right) .
\end{equation}%
Using these expressions the Dyson map is obtained from (\ref{eta}) as%
\begin{equation}
\eta _{1}=\eta _{4}=\sqrt{\xi },~~\eta _{2}=\frac{\hat{\xi}}{\sqrt{\xi }}%
,~~\eta _{3}=\frac{1}{\sqrt{\xi }}-\sqrt{\xi },~~~~~\text{with }\hat{\xi}%
:=A\mu +\sqrt{AB-1}.
\end{equation}%
In this case we compute $\det \eta =\eta _{1}^{2}-\eta _{2}^{2}-\eta
_{3}^{2}=\pm 2\delta $ with $\delta =A-1$ .

The energy operator (\ref{Henergy}) acquires the form 
\begin{equation}
\tilde{H}(t)=-\frac{1}{2}\left\{ \omega \mathbb{I}+\frac{\chi }{\delta }%
\left[ i\left( \xi -1\right) \sigma _{x}+i\hat{\xi}\sigma _{y}+(\xi -\delta
)\sigma _{z}\right] \right\}
\end{equation}%
Similarly as above we construct the antilinear symmetry operator for this
operator 
\begin{equation}
\widetilde{\mathcal{PT}}:=\frac{1}{\sqrt{(\xi -\delta )^{2}+\hat{\xi}^{2}}}%
\left[ i\hat{\xi}\sigma _{y}+(\xi -\delta )\sigma _{z}\right] \tau .
\label{PTex}
\end{equation}%
The eigenvectors of $\tilde{H}(t)$ are computed to%
\begin{equation}
\tilde{\varphi}_{\pm }\sim \left( 
\begin{array}{c}
(1\mp 1)\delta -\xi \\ 
\hat{\xi}+i(1-\xi )%
\end{array}%
\right) ,
\end{equation}%
which are indeed $\widetilde{\mathcal{PT}}$-eigenstates, that is we have $%
\widetilde{\mathcal{PT}}\tilde{\varphi}_{\pm }=e^{i\tilde{\omega}_{\pm }}%
\tilde{\varphi}_{\pm }$ with 
\begin{eqnarray}
\tilde{\omega}_{+} &=&\arctan \left[ \frac{(1-\xi )\hat{\xi}}{1-(1+A)\xi
+\xi ^{2}}\right] +\pi , \\
\tilde{\omega}_{-} &=&\arctan \left[ \frac{\sqrt{1-\alpha ^{2}}(1-\alpha \xi
)\hat{\xi}}{3+2A(A-2)-(3-A)\xi +\xi ^{2}}\right] .
\end{eqnarray}%
This means as long as $AB>1$ the energy operator $\tilde{H}(t)$ is $%
\widetilde{\mathcal{PT}}$-symmetric with regard to (\ref{PTex}).

\section{Conclusions}

We have demonstrated that a non-Hermitian Hamiltonian in its spontaneously
broken $\mathcal{PT}$-symmetric phase allows for a self-consistent quantum
mechanical description when an explicit time-dependence is introduced into
its parameters. This is possible as the Hamiltonian that satisfies the
time-dependent Schr\"{o}dinger equation becomes unobservable and instead the
energy operator develops real eigenvalues at any instance in time. We
identified the new antilinear operator $\widetilde{\mathcal{PT}}$ that
explains the reality of the spectrum of the energy operator in parts of the
parameter regime.

We have solved for the first time the time-dependent Dyson equation in
conjunction with the time-dependent Schr\"{o}dinger equation in complete
generality. Previously only special cases were considered, e.g. one of the
Hamiltonians was kept time-independent or just the time-dependent Dyson
equation was solved without further elaboration on whether the solutions
obtained can be used in the solutions to the time-dependent Schr\"{o}dinger
equation.

Naturally it would be interesting to investigate different types of models
and in particular extend the analysis to systems with infinite dimensional
Hilbert spaces.

\bigskip \noindent \textbf{Acknowledgments:} TF is supported by a City,
University of London Research Fellowship.

\newif\ifabfull\abfulltrue

%%\bibliographystyle{phreport}
%%\bibliography{acompat,Ref}

\begin{thebibliography}{99}
\bibitem{EW} E.~Wigner, \newblock Normal form of antiunitary operators, %
\newblock J. Math. Phys. \textbf{1}, 409--413 (1960).

\bibitem{Bender:1998ke} C.~M. Bender and S.~Boettcher, \newblock Real
Spectra in Non-Hermitian Hamiltonians Having PT Symmetry, \newblock Phys.
Rev. Lett. \textbf{80}, 5243--5246 (1998).

\bibitem{BFGJ} C.~M. Bender, A.~Fring, U.~Guenther, and H.~F. Jones, %
\newblock Special issue on quantum physics with non-Hermitian operators, %
\newblock J. Phys. A: Math. and Theor. \textbf{45}(1), 010201 (2012).

\bibitem{Urubu} F.~G. Scholtz, H.~B. Geyer, and F.~Hahne, \newblock %
Quasi-Hermitian Operators in Quantum Mechanics and the Variational
Principle, \newblock Ann. Phys. \textbf{213}, 74--101 (1992).

\bibitem{Benderrev} C.~M. Bender, \newblock Making sense of non-Hermitian
Hamiltonians, \newblock Rept. Prog. Phys. \textbf{70}, 947--1018 (2007).

\bibitem{Alirev} A.~Mostafazadeh, \newblock Pseudo-Hermitian Representation
of Quantum Mechanics, \newblock Int. J. Geom. Meth. Mod. Phys. \textbf{7},
1191--1306 (2010).

\bibitem{siegl2012metric} P.~Siegl and D.~Krej{\v{c}}i{\v{r}}{\'\i}k, %
\newblock On the metric operator for the imaginary cubic oscillator, %
\newblock Phys. Rev. D \textbf{86}(12), 121702 (2012).

\bibitem{bagarello2013self} F.~Bagarello, \newblock From self-adjoint to
non-self-adjoint harmonic oscillators: Physical consequences and
mathematical pitfalls, \newblock Phys. Rev. A \textbf{88}(3), 032120 (2013).

\bibitem{FabFring1} F.~Bagarello and A.~Fring, \newblock Non-self-adjoint
model of a two-dimensional noncommutative space with an unbound metric, %
\newblock Phys. Rev. A \textbf{88}(4), 042119 (2013).

\bibitem{krejvcivrik2015pseudospectra} D.~Krej{\v{c}}i{\v{r}}{\'\i}k,
P.~Siegl, M.~Tater, and J.~Viola, \newblock Pseudospectra in non-Hermitian
quantum mechanics, \newblock J. of Math. Phys. \textbf{56}(10), 103513
(2015).

\bibitem{Muss} Z.~H. Musslimani, K.~G. Makris, R.~El-Ganainy, and D.~N.
Christodoulides, \newblock Optical Solitons in PT Periodic Potentials, %
\newblock Phys. Rev. Lett. \textbf{100}, 030402 (2008).

\bibitem{MatMakris} K.~G. Makris, R.~El-Ganainy, D.~N. Christodoulides, and
Z.~H. Musslimani, \newblock PT-symmetric optical lattices, \newblock Phys.
Rev. \textbf{A81}, 063807(10) (2010).

\bibitem{Guo} A.~Guo, G.~J. Salamo, D.~Duchesne, R.~Morandotti,
M.~Volatier-Ravat, V.~Aimez, G.~A. Siviloglou, and D.~Christodoulides, %
\newblock Observation of PT-Symmetry Breaking in Complex Optical Potentials, %
\newblock Phys. Rev. Lett. \textbf{103}, 093902(4) (2009).

\bibitem{CA} C.~Figueira~de Morisson~Faria and A.~Fring, \newblock Time
evolution of non-Hermitian Hamiltonian systems, \newblock J. Phys. A \textbf{%
39}, 9269--9289 (2006).

\bibitem{CArev} C.~Figueira~de Morisson~Faria and A.~Fring, \newblock %
Non-Hermitian Hamiltonians with real eigenvalues coupled to electric fields:
from the time-independent to the time dependent quantum mechanical
formulation, \newblock Laser Physics \textbf{17}, 424--437 (2007).

\bibitem{fringmoussa} A.~Fring and M.~H.~Y. Moussa, \newblock Unitary
quantum evolution for time-dependent quasi-Hermitian systems with
nonobservable Hamiltonians, \newblock Phys. Rev. A \textbf{93}(4), 042114
(2016).

\bibitem{fringmoussa2} A.~Fring and ~M. H.~Y. Moussa, \newblock %
Non-Hermitian Swanson model with a time-dependent metric, \newblock Phys.
Rev. A \textbf{94}(4), 042128 (2016).

\bibitem{AndTom1} A.~Fring and T.~Frith, \newblock Exact analytical
solutions for time-dependent Hermitian Hamiltonian systems from static
unobservable non-Hermitian Hamiltonians, \newblock Phys. Rev. A \textbf{95},
010102(R) (2017).

\bibitem{AndTom2} A.~Fring and T.~Frith, \newblock Metric versus observable
operator representation, higher spin models, \newblock preprint
arXiv:1612.06122 (2016).

\bibitem{time1} A.~Mostafazadeh, \newblock Time-dependent pseudo-Hermitian
Hamiltonians defining a unitary quantum system and uniqueness of the metric
operator, \newblock Phys. Lett. B \textbf{650}(2), 208--212 (2007).

\bibitem{time6} M.~Znojil, \newblock Time-dependent version of
crypto-Hermitian quantum theory, \newblock Phys. Rev. D \textbf{78}(8),
085003 (2008).

\bibitem{time7} J.~Gong and Q.-H. Wang, \newblock Time-dependent
PT-symmetric quantum mechanics, \newblock J. Phys. A: Math. and Theor. 
\textbf{46}(48), 485302 (2013).

\bibitem{Ermakov} V.~Ermakov, \newblock Transformation of differential
equations,, \newblock Univ. Izv. Kiev. \textbf{20}, 1--19 (1880).

\bibitem{Pinney} E.~Pinney, \newblock The nonlinear differential equation $%
y^{\prime \prime 3}=0$, \newblock Proc. Amer. Math. Soc. \textbf{1}, 681(1)
(1950).

\bibitem{leach2008ermakov} P.~G.~L. Leach and K.~Andriopoulos, \newblock The
Ermakov equation: a commentary, \newblock Applicable Analysis and Discrete
Mathematics , 146--157 (2008).
\end{thebibliography}

\end{document}